\documentclass{aa}

\usepackage{graphicx}
\usepackage{natbib}
\usepackage{amsmath}
\usepackage[breaklinks=true,colorlinks=true,urlcolor=blue,citecolor=blue,linkcolor=blue]{hyperref} %% to avoid \citeads line fills

\newcommand{\be}{\begin{eqnarray}}
\newcommand{\ee}{\end{eqnarray}}

\def\HeI{\ion{He}{I}}
\def\HeII{\ion{He}{II}}

\def\Hefef{\HeI~\ensuremath{\lambda 584}}
\def\Hetof{\HeII~\ensuremath{\lambda 304}}
\def\Hetfs{\HeII~\ensuremath{\lambda 256}}

\def\dd{\mathrm{d}}
\def\dnu{\ensuremath{\dd \nu}}

\newcommand{\mean}[1] {\langle #1 \rangle}

\begin{document}
\title{Formation of the helium EUV resonance lines}

\author{T. P. Golding \inst{1} \and 
  J. Leenaarts \inst{2} \and
  M. Carlsson \inst{1}}

\institute{Institute of Theoretical Astrophysics, University of Oslo,
    P.O. Box 1029 Blindern, NO-0315 Oslo, Norway \\
    \email{thomas.golding@astro.uio.no, mats.carlsson@astro.uio.no}
    \and
    Institute for Solar Physics, Department of Astronomy, Stockholm
    University, AlbaNova University Centre, SE-106 91 Stockholm,
    Sweden \\ \email{jorrit.leenaarts@astro.su.se}}

\abstract
{While classical models 
successfully reproduce intensities 
of many transition region lines, they predict helium EUV line 
intensities roughly an order of magnitude lower than the observed value.}
{To determine the relevant formation mechanism(s) of the helium EUV resonance lines,
capable of explaining the high intensities under quiet sun conditions.}
{We synthesise and study the emergent spectra from a 3D radiation-magnetohydrodynamics 
simulation model. The effects of coronal illumination and non-equilibrium ionisation of hydrogen 
and helium are included self-consistently in the numerical simulation.}
{Radiative transfer calculations result in helium EUV line intensities that are an order 
of magnitude larger than the intensities calculated under the classical assumptions. The 
enhanced intensity of \Hefef\ is primarily caused by \HeII\ recombination cascades. The enhanced
intensity of \Hetof\ and \Hetfs\ is caused  primarily by non-equilibrium helium ionisation.}
{The analysis shows that the long standing problem of the high helium EUV line intensities disappears
when taking into account optically thick radiative transfer and non-equilibrium ionisation effects.}

%\keywords{radiative transfer --- Sun: atmosphere --- Sun: chromosphere}

\maketitle

\section{Introduction}
The formation mechanism of the \ion{He}{i} and \ion{He}{ii} EUV resonance lines is a topic that has been 
discussed for many decades.  
Some of the first calibrated spectral observations of the 
EUV helium lines, and the EUV wavelength range in general,
became available in the sixties around the time when the
Orbiting Solar Observatory series of satellites were launched. 
While models were capable of explaining the intensities of many transition region lines,
they did not explain the helium intensities that were observed.
In particular, 
\citet{1975MNRAS.170..429J}
found that the observed intensity of these 
lines were an order of magnitude larger than derived values
based on emission measure models. 
Values in similar
ranges have been confirmed also by more recent studies
\citep{1999MNRAS.308..510M, pietarila2004, 2015ApJ...803...66G}
using classical modelling 
similar to that of 
\citet{1975MNRAS.170..429J}.
Pinning down the relevant formation 
mechanism(s) of the helium lines is important because it will help 
guide interpretations of observations.

A possible intensity enhancement mechanism 
was suggested already by 
\citet{1975MNRAS.170..429J}. 
The collisional excitation rate of the 
helium resonance lines are highly sensitive to temperature. She therefore proposed that if
cold ions in some way got mixed with hot electrons, more photons would be produced.
One possible physical mechanism that would create such a mix is particle diffusion.
This was added by \cite{1990ApJ...355..700F, 1991ApJ...377..712F, falc1993} in a 
semi-empirical 1D model, and gave enough intensity
in the hydrogen Lyman-$\alpha$ without the need to for the temperature plateau at 20\,kK 
present in the \cite{val3c1981} models.
Another mechanism is the so called velocity redistribution (VR) \citep{jordan1980}
which was also investigated by 
\cite{andretta2000} who invented the term VR,  \cite{2002MNRAS.337..666S}, and 
\cite{pietarila2004}.
The driver of the latter  mechanism is  turbulent fluid motions that 
transport cold atoms into hotter regions. 
While both the particle diffusion and the VR mechanisms do enhance the helium EUV line 
intensities, they affect only the photon production due to collisional excitation.

Another suggested mechanism is the photoionisation-recombination (PR) mechanism
\citep{zirin1975,1996SoPh..169..313Z}. 
In short; half of the EUV photons emitted in transition region
and coronal lines will be lost into space and the other half will constitute a coronal 
illumination of the chromosphere and  be absorbed in the continua of either helium or hydrogen. The resulting 
photoionised helium atoms will recombine and de-excite through a cascade event, and ultimately 
emit a helium resonance photon. 
In such a scenario more EUV photons emitted from the corona would increase 
the number of helium resonance line photons produced. 
\citet{andretta1997} 
investigated this idea 
and found that incident radiation from the corona could enhance the line intensity 
of \ion{He}{i} $\lambda$584 in very quiet regions, but had a marginal effect on more 
active regions. Later he also found that the PR mechanism can not
be the primary formation mechanism of \ion{He}{ii} $\lambda$304 in quiet regions 
\citep{andretta2003}.
The subordinate helium lines, however, have been shown to be sensitive to coronal illumination
\citep{wahlstrom1994,avrett1994, andretta1997, mauas2005, centeno2008, leenaarts2016}.

In this paper we study the formation of the helium EUV resonance lines by exploiting 
state-of-the-art 3D numerical models and 3D non-LTE radiative transfer including the effects
of non-equilibrium ionisation.
The paper is laid out as follows: in Sec.~\ref{section:method} we describe our method of computing the line intensities, in Sec.~\ref{section:atommodel}--\ref{section:modelatmosphere} we describe our model atom and 
model atmosphere, in Sec.~\ref{section:classic} we follow a classical approach and assess
the helium line enhancement calculated from our model,  in Sec.~\ref{section:rteffects} we describe how the helium resonance lines are formed in the model, 
and in Sec.~\ref{section:discussion} we discuss our results. Finally, in Sec.~\ref{section:summary}
we summarise and draw conclusions.

%%%%%%%%%%%%%%%%%%%%
\section{Method}\label{section:method}
%%%%%%%%%%%%%%%%%%%%
We compare and investigate line intensities from the solar atmosphere
at disc centre ($\mu = 1$). 
The line intensity, $I$, is given by the frequency integral over the specific intensity 
emerging from the atmosphere, $I_\nu(\mu)$, 
\begin{eqnarray}
  I = \int_D I_\nu(\mu=1) \, \dnu \label{eq:lineint},
\end{eqnarray}
where $\nu$ is frequency and $D$ denotes the frequency domain 
relevant for the line under consideration.
$I_\nu(\mu)$ is the solution of the equation of radiative transfer
\begin{eqnarray}
 \frac{\dd I_\nu}{\dd s} = \eta_\nu - I_\nu \chi_\nu \label{eq:rt},
\end{eqnarray}
where $s$ is the distance along a ray, $\eta_\nu$  and $\chi_\nu$ are the frequency 
dependent emissivity and opacity, respectively. In Sec.~\ref{section:modified_multi} and~\ref{section:thinformation} we describe how we solve Eq.~\ref{eq:rt}.

 %%%%%%%%%%%%%%%%%%%%%%%%%%%%%%
\subsection{Non-equilibrium radiative transfer with Multi3d}
\label{section:modified_multi}
%%%%%%%%%%%%%%%%%%%%%%%%%%%%%%

We use Multi3d 
\citep{leenaarts2009} 
to solve the equation of radiative transfer
coupled to a particle conservation equation and a set of equilibrium rate equations.
The transfer equation is solved using an extremum-preserving third order Hermite interpolation scheme 
\citep[e.g.][]{2013A&A...549A.126I}. 
The code solves the rate equations using the formulation of 
 \citet{1991A&A...245..171R, 1992A&A...262..209R} 

Given a model atom with $N$ atomic states constituting $M$ ion stages and a 
set of transitions, the statistical equilibrium rate equations take the form
\begin{eqnarray}
 \sum_{j=1,j \ne i}^N n_j P_{ji} - n_i \sum_{j=1, j\ne i}^{N}   P_{ij} = 0, \label{eq:rate}
\end{eqnarray}
where $P_{ij}$ is the rate coefficient giving the probability per unit time of a transition from the
$i$-th to the $j$-th state. In principle it is possible to express $N$ equilibrium rate equations, but
they constitute a linearly dependent set of equations which is why we replace one of them by the particle 
conservation equation,
\begin{eqnarray}
 \sum_{i=1}^N n_i = n, \label{eq:partcons}
\end{eqnarray}
where $n$ is proportional to the mass density by an abundance dependent 
factor. We refer to the intensity calculated with this setup, solving the radiative transfer equation 
(Eq. \ref{eq:rt}) together with $N-1$ rate equations (Eq. \ref{eq:rate}) and one particle 
conservation equation (Eq. \ref{eq:partcons}), as the non-LTE solution (NLTE).

We know that non-equilibrium effects on the helium lines are 
lost in the NLTE solution 
\citep{golding2014}. 
In that paper we also showed that the processes that 
affect the ionisation state of helium are slow compared to the processes that populate
the excited states. We can therefore approximate non-equilibrium effects by constraining the 
radiative transfer solution by non-equilibrium ion fractions
\citep[for an analysis of the time-dependent rate equations in evolving plasmas see][]{2005JQSRT..92..479J}.
We do this by solving the equilibrium rate
equations (Eq.~\ref{eq:rate}) for the $N-M$ excited states of the atom. The $M$ ion fraction equations replace
the ground state rate equations. Let $f_j$ denote the fraction of atoms
in the $j$-th ion stage. The ion fraction equations then take the form
 \begin{eqnarray}
 \sum_{i=1}^N n_i d_{ji} = f_j n \label{eq:ionstage}
\end{eqnarray}
where $d_{ij}=1$ if $i$ is a state in the $j$-th ion stage and $d_{ij}=0$ if $i$ is not a state in
the $j$-th ion stage. We modify Multi3d so that it solves this alternative set
of equations when ion fractions are given as input. We refer to the solution of this
set of equations as the non-equilibrium non-LTE solution (NE-NLTE). 

%

%%%%%%%%%%%%%%%%%%%
\subsubsection{Test of the non-equilibrium non-LTE radiative transfer}
%%%%%%%%%%%%%%%%%%%

%%%%%%%
\begin{figure*}
 \includegraphics[width=\textwidth]{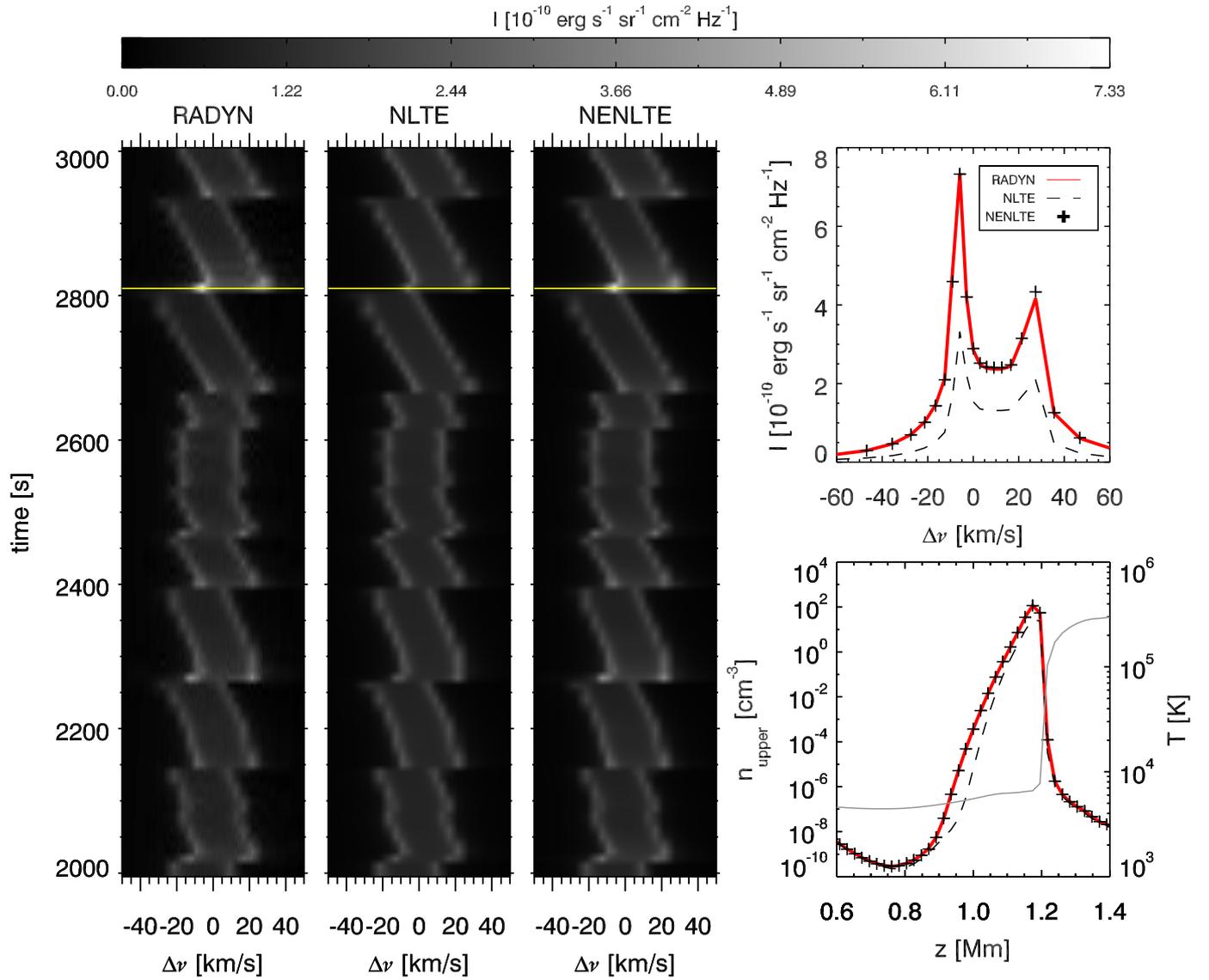}
   \caption{Emergent \Hefef\ line profiles from the Radyn simulation 
   compared to the NLTE and NE-NLTE radiative transfer solutions from Multi 
   (three left panels). The top right panel shows the line profile 
   in the three approaches for the snapshot indicated by the yellow 
   lines. The bottom right panel shows the number density of the upper 
   level of the transition in the three approaches. The grey line shows
   the temperature with scale to the right.}
 \label{fig:584}
\end{figure*}
%%%%%%

We now check whether the method described in Sec.~\ref{section:modified_multi} 
for computing the NE-NLTE solution matches the emergent intensity of a dynamic 
solar atmosphere model. Such a test can only be carried out in a 1D geometry because it
is not possible with contemporary computer resources to construct a dynamic 3D model with 
a self-consistent time dependent non-LTE description of the radiative transfer.

The dynamic solar atmosphere model is computed with the radiation-hydrodynamics code
Radyn 
\citep{carlsson_stein1992,carlsson_stein1995,carlsson_stein1997,carlsson_stein2002}. 
Radyn solves the conservation equations of mass, momentum, 
charge, and energy, as well as the radiative transfer equation (Eq. \ref{eq:rt}) and a set of
non-equilibrium rate equations for the atomic number densitites. 
\begin{eqnarray}
\frac{\dd n_i}{\dd t}  + \nabla \cdot (\mathbf{v}n_i) =  
\sum_{j=1,j \ne i}^N n_j P_{ji} - n_i \sum_{j=1, j\ne i}^{N}   P_{ij},
\label{eq:nerate}
\end{eqnarray}
where $\mathbf{v}$ is the bulk velocity.

We use the NE-run from 
\cite{golding2014} 
as a reference model. Each snapshot 
of the reference model describes the state of the atmosphere at a given time.
For each snapshot we write input files for the 1D radiative transfer code Multi
\citep{carlsson:multi} 
and compute the NLTE solution. The modifications made to 
Multi3d are made also in Multi. Non-equilibrium ion fractions, $f_j$, are available 
from the reference Radyn model. We use these as input together with the Multi 
input files and for each snapshot compute the NE-NLTE solution with the modified version of Multi.

In Fig.~\ref{fig:584} we compare the time dependent emergent intensity of 
the \Hefef\ resonance line with the results from the NLTE and NE-NLTE solutions.
The sawtooth pattern due to shocks is evident in all three intensities. The NE-NLTE solution  
reproduces the intensity brightenings occurring at the onset of some of shocks.
This is a non-equilibrium effect and it is not recovered in the NLTE solution.
The \Hetof\ line has  brightenings as well (shown in Figure 8 of \cite{golding2014}),
and also these are reproduced in the NE-NLTE solution and lost in the NLTE solution (not 
shown here).

From this we conclude that the NE-NLTE method implemented in Multi3d will 
produce relevant non-equilibrium effects on the helium spectrum, given the appropriate 
non-equilibrium ionisation state as input.

%%%%%%%%%%%%%%%%
\subsection{Optically thin line formation}
\label{section:thinformation}
%%%%%%%%%%%%%%%%

Assuming optically thin conditions we neglect the last term of Eq.~\ref{eq:rt}. 
In this case the line intensity becomes a simple line-of-sight integral along the ray $S$,
\begin{eqnarray}
  I_\mathrm{th} = \int_{S} \eta \ \dd s \label{eq:thinint},
\end{eqnarray}
where $\eta$ is the frequency-integrated line emissivity given by
$\eta = \int_D \eta_\nu \ d\nu$. The frequency integrated emissivity can be expressed as
\begin{eqnarray}
    \eta=A_{ul}\left(\frac{h\nu_0}{4\pi}\right)
  \left(\frac{n_{u}}{n_\mathrm{ion}} \right)
  \left(\frac{n_\mathrm{ion}}{n_\mathrm{el}}\right) \left(
  \frac{n_{\mathrm{el}}}{n_{\mathrm{H}}} \right) n_\mathrm{H},
  \label{eq:emissivity}
\end{eqnarray} 
where $A_{ul}$ is the Einstein coefficient for spontaneous radiative
de-excitation, $h$ is Planck's constant, $\nu_0$ is the frequency of the
line, $n_u$ is the number density of atoms in the upper level of the line,
$n_\mathrm{ion}$ is the number density of atoms in the ion stage,
$n_\mathrm{el}$ is the number density of atoms of the element, and
$n_{\mathrm{H}}$ is the number density of hydrogen atoms. This expression 
is general; $n_u$ is dependent on the intensity, as well as 
$n_\mathrm{e}$ and $T$. However, under optically thin conditions $n_u$ looses its
dependency on the intensity. If we in addition assume that
\begin{itemize}
\item the processes that are affecting  $n_\mathrm{ion}$ are independent of 
the processes affecting the number densities of the excited states within
an ion stage,
\item the system is in a steady state condition,
\end{itemize}
then the evaluation of Eq.~\ref{eq:emissivity} becomes simple. We will refer to these 
three assumptions collectively as the optically thin equilibrium approximation. The fraction
$(n_u/n_\mathrm{ion})$ is then found by solving a set of 
equilibrium rate equations (Eq. \ref{eq:rate}) and a normalised particle conservation equation
(Eq. \ref{eq:ionstage} divided by $n_\mathrm{ion}$). The transition rate coefficients are only 
made up of terms involving collisional excitation, collisional de-excitation and spontaneous 
radiative de-excitation. The collisional terms are expressed by $n_\mathrm{e}q(T)$, where $q(T)$
is dependent on the specific transition. The radiative spontaneous de-excitation rate is given by  the Einstein
$A_{ul}$ coefficient. 
The ionisation equilibrium fraction, $(n_\mathrm{ion}/n_\mathrm{el})$, is found in a similar way. 
The transition rate coefficients are in this case made up of terms made up of collisional ionisation, di-electronic 
recombination and spontaneous radiative recombination. All of these terms are linear in 
$n_\mathrm{e}$, and this makes $(n_\mathrm{ion}/n_\mathrm{el})$ dependent on temperature
only.
Under the optically thin equilibrium approximation we denote these two fraction as 
$(n_\mathrm{u}/n_\mathrm{ion})_\mathrm{eq}$ and $(n_\mathrm{ion}/n_\mathrm{el})_\mathrm{eq}$. 
The optically thin equilibrium emissivity, $\eta_\mathrm{eq}$, is then expressed as
\begin{eqnarray}
  \eta_\mathrm{eq} = G(T,n_\mathrm{e}) \, n_\mathrm{e} n_\mathrm{H}, \label{eq:emiss_eq}
\end{eqnarray}
where $G(T,n_\mathrm{e})$ is defined as
\begin{eqnarray}
  G(T,n_\mathrm{e}) = \left( \frac{A_{ul}}{n_\mathrm{e}}\right) \left(\frac{h\nu_0}{4\pi}\right)
  \left(\frac{n_{u}}{n_\mathrm{ion}} \right)_\mathrm{eq}
  \left(\frac{n_\mathrm{ion}}{n_\mathrm{el}}\right)_\mathrm{eq} \left(
  \frac{n_{\mathrm{el}}}{n_{\mathrm{H}}} \right).
\end{eqnarray}
This function will typically have  only a weak dependence on $n_\mathrm{e}$ and be sharply peaked in 
temperature. We will refer to the
temperature where $G$ peaks as the formation temperature of the line.

\subsubsection{Differential emission measure modelling}
A differential emission measure distribution (DEM), $\Phi(T)$, gives an indication of how much
radiating material is present at different temperatures of a solar region or feature. It is
defined as,
\begin{eqnarray}
  \Phi(T) = n_{\mathrm{H}} n_\mathrm{e} \frac{\dd z}{\dd T}. 
\end{eqnarray}
It can be used to compute the thin line intensities under equilibrium conditions,
\begin{eqnarray}
  I_\mathrm{DEM} = \int_T G(T,n_\mathrm{e}) 
                        \Phi(T) \ \dd T \label{eq:dem}
\end{eqnarray}
When a set of line intensities are known, it is possible to invert this equation and obtain 
an estimate of $\Phi(T)$ 
\citep[see for example][and references therein]{2015ApJ...807..143C}.
Since $G$ is also a (weak) function of $n_\mathrm{e}$ this
requires an assumption about the electron density as function of temperature. A common assumption is to use a  constant electron pressure, 
$P_\mathrm{e} = k_\mathrm{B} n_\mathrm{e}T$, where $k_\mathrm{B}$ is Boltzmann's
constant.

%%%%%%%%%%%%%% 
\section{Helium model atom}\label{section:atommodel}
%%%%%%%%%%%%%%

We use a 13 level helium model atom to perform radiative transfer calculations. It has 
nine \ion{He}{i} states, three \ion{He}{ii} states and the fully ionised state \ion{He}{iii}. To construct
the 13 level model atom we used the 33 level helium model atom from \cite{golding2014} as a 
basis. For \ion{He}{i} we kept the ground state $n=1$ and six $n=2$ excited states unchanged. All
singlet $n=3$ states were merged to one representative state. This was also done 
for all triplet $n=3$ states. All the \ion{He}{i} excited states with $n\ge 4$ were neglected.
For \ion{He}{ii} we kept the ground state $n=1$ unchanged. All the $n=2$ states were merged to 
one representative state, and all the $n=3$ states were merged to one representative state. \ion{He}{ii}
excited states with $n\ge 4$ were neglected. The helium abundance is set to 0.1 helium atoms
per hydrogen atom.

%%%%%%%%%%%%%%
\section{Model atmosphere}\label{section:modelatmosphere}
%%%%%%%%%%%%%%

As our model atmosphere we use a snapshot from a 3D radiation-magnetohydrodynamic 
simulation. The simulation was run with the code Bifrost 
\citep{gudiksen2011} 
and features 
non-equilibrium ionisation of hydrogen 
\citep{leenaarts2007}
 and helium 
 \citep{golding2016}.
The ionisation state of helium is strongly dependent on the radiative losses from the transition 
region and corona (EUV photons) which leads to photoionisation in the chromosphere. 
The radiative losses are taken into account in the simulation and they self-consistently give 
rise to the coronal illumination. 
Other than the non-equilibrium helium ionisation the simulation has the same setup as the
enhanced network simulation described in 
\citet{2016A&A...585A...4C}. 
The spatial domain of the simulation is $24\times 24\times 17$ Mm$^3$, spanning from the convection 
zone at $z=-2.5$ Mm to the corona at $z=14.5$ Mm. 
$z=0$ Mm is defined as the average height where the the optical depth at 5000 \AA\ is unity.
The simulation has a resolution of $504\times 504 \times 496$ grid points. 
The snapshot we use as the model atmosphere represents the physical state of the atmosphere 
about 20 minutes after the non-equilibrium ionisation was switched on. This is long 
enough for potential startup effects to have vanished. According to figures shown in 
\cite{golding2016}, 
the ionisation state of helium relaxes to chromospheric conditions
on timescales of about 10-15 minutes. 
The simulation provides all the quantities that we need ($\rho$, $\mathbf{v}$, $T$, $n_\mathrm{e}$, $n_i$ of hydrogen, and
$f_j$ of helium) at each grid point of the snapshot.

%%%%%%%%%%%%
\section{Enhancement factors and the classical approach}
\label{section:classic}
%%%%%%%%%%%%
%%%%%%%%
\begin{table*}
\caption{Helium line enhancement factors from different studies.}
\label{table:efactors}
\centering
\begin{tabular}{cccc}
\hline \hline
    Study & \ion{He}{i} $\lambda$584  & 
                  \ion{He}{ii} $\lambda$304 & \ion{He}{ii} $\lambda$256 \\ \hline
  \cite{1975MNRAS.170..429J} & 15 & 5.5 &  \\
  \cite{1999MNRAS.308..510M} & 10-14 & 13-25 &   \\
  \cite{pietarila2004}  & 2-10 & 27 & \\
  \cite{2015ApJ...803...66G} & 0.5-2 & 13 & 5 \\ 
  This work &  7 & 10 & 7 \\
\hline
\end{tabular}
\end{table*}
%%%%%%

We will follow a similar procedure to that of  
\cite{2015ApJ...803...66G}.
In short, they observed a set of lines with formation temperatures in the range
$^{10}$log$(T)=4.3$ to 6.25, including resonance lines of \ion{He}{i} and \ion{He}{ii}.
Line intensities were spatially and temporally averaged. Based on a subset
of these line intensities they inverted Eq.~\ref{eq:dem} to obtain a DEM. Then they used the derived
DEM to compute the intensities of all the observed lines, $I_\mathrm{DEM}$ (also with Eq.~\ref{eq:dem}).
The discrepancy between observed and modelled line intensities was quantified by an enhancement factor
defined as the ratio of the observed intensity to the DEM-modelled intensity. For most of the lines the enhancement was around one,
except for a few lines including the resonance lines of \ion{He}{ii}.
Similar studies have have been conducted also by others \citep{1975MNRAS.170..429J, 1999MNRAS.308..510M, 
pietarila2004}. In Table \ref{table:efactors} we sum
up their resulting enhancement factors for the helium lines.

The lines observed by \cite{2015ApJ...803...66G} are listed in their Table 2, 3, and 4. We compute the
line intensities for all of the listed lines assuming optically thin equilibrium conditions (defined in Section
\ref{section:thinformation}). For \Hefef, \Hetof, and \Hetfs\ we also compute the
NE-NLTE thick line intensities.

\subsection{Thin calculations}
We assume optically thin equilibrium conditions and compute the EUV line intensities with 
Eq.~\ref{eq:thinint} from all the columns of the model atmosphere. 
To obtain the equilibrium line emissivities, $\eta_\mathrm{eq}$, of the various lines on each point
of the model atmosphere we exploit the atomic database and software tools from the Chianti package 
\citep{dere1997, 2015A&A...582A..56D}. We use the \texttt{chianti.ioneq} ionisation equilibrium 
file and coronal abundances from \cite{2012ApJ...755...33S}. For \Hefef, \Hetof, and \Hetfs\ we
use the helium model atom described in Section \ref{section:atommodel} to calculate $\eta_\mathrm{eq}$. 
These thin intensities are denoted $I_\mathrm{th}$.

\subsection{Thick calculations}
We  run the NE-NLTE mode of Multi3d to compute the non-equilibrium helium line intensities 
from the model atmosphere. To make the problem computationally tractable we halved the 
horizontal resolution of the model atmosphere so that it is given on a $252\times 252 \times 496$ 
grid. The line intensities computed are \ion{He}{i} $\lambda$584, \ion{He}{ii} $\lambda$304, 
and the \ion{He}{ii} $\lambda$256. These line intensities are denoted $I_\mathrm{RT}$.

%%%%%%%%%%%%%%%%%%%%%
\subsection{Verification of the model atmosphere}\label{section:verification}
%%%%%%%%%%%%%%%%%%%%%%

%%%%%%%%
\begin{figure*}
 \includegraphics[width=\textwidth]{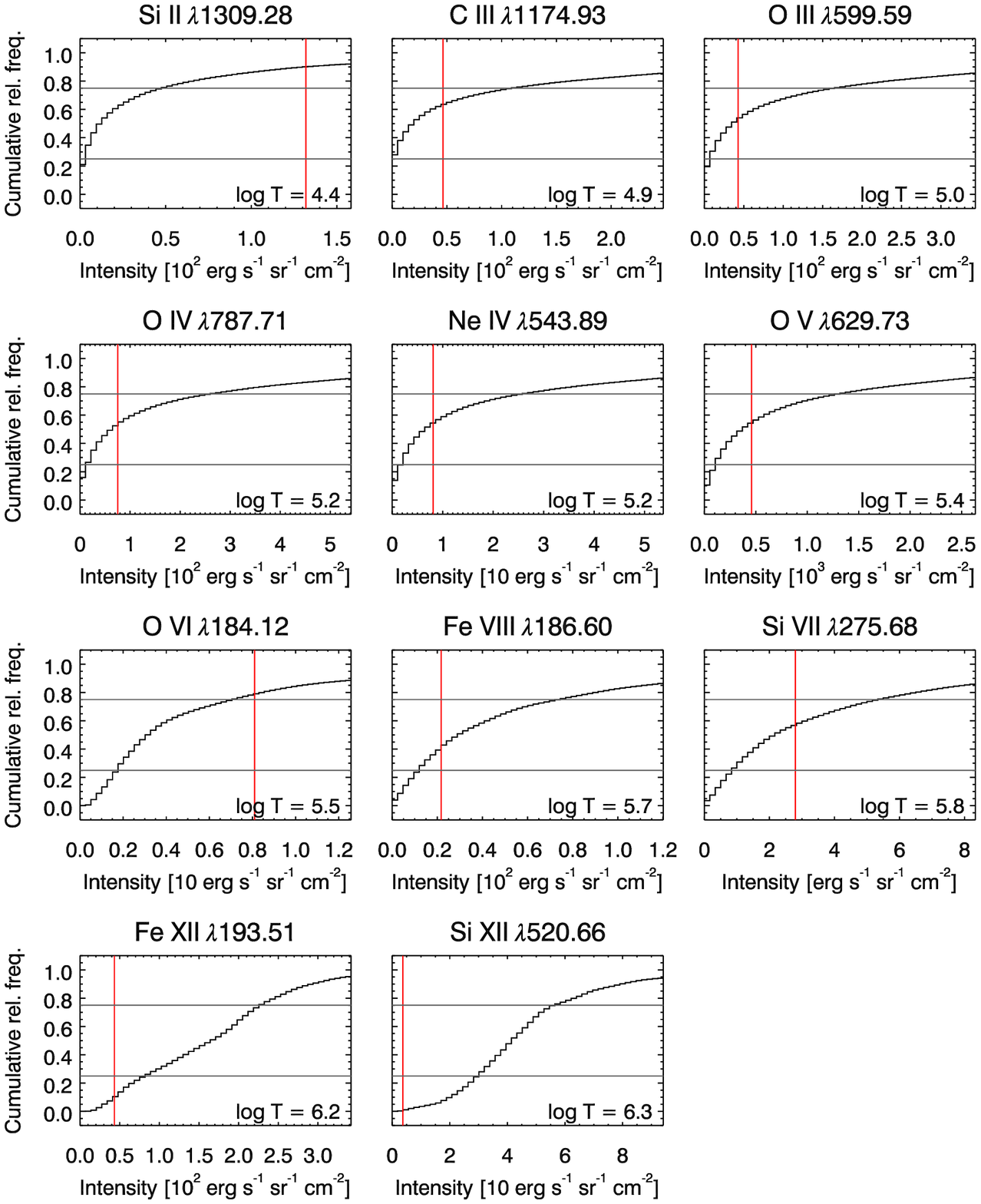}
   \caption{Cumulative distributions of thin line intensities, $I_\mathrm{th}$,
    computed from the model
   atmosphere. Horizontal lines in each panel indicate the
   location of the first and third quartile. Red vertical lines indicate the
   observed values.}
 \label{fig:thin_giunta_comparison}
\end{figure*}
%%%%%%%%

To assess how well our model atmosphere represents solar transition region
and coronal conditions we compare our intensity, $I_\mathrm{th}$, 
with observed values. The selection of lines and their observed values are listed
in Table 7 of \cite{2015ApJ...803...66G}. These are the lines they used to derive their
DEM distribution, and they are selected to span a large temperature range. The observed
values are of a quiet sun region.

Figure \ref{fig:thin_giunta_comparison} shows the 
cumulative distribution of $I_\mathrm{th}$ for the selected lines.
The red vertical line in each panel indicates the observed value, and the horizontal grey
lines show the first and third quartile. Our model seem to produce reasonable 
intensities compared to the observations in most of the lines shown. There are exceptions 
in the low and high temperature ends. The synthetic intensities of 
the low temperature \ion{Si}{ii} $\lambda$1309 line are likely not very realistic since this line
does not form under optical thin
conditions \citep{1994A&A...287..972L}. The observed intensity of the high-temperature \ion{Si}{xii} $\lambda$520 line is far below the interquartile range of the 
synthetic intensity distribution.
However, since the model atmosphere represents an enhanced network region, we
expect there to be more hot plasma present than what is the case for quiet sun conditions, 
and thus higher synthetic intensities of the hot lines.

The model atmosphere produces transition region and 
coronal line intensities of the same order of magnitude as what is observed in quiet sun regions. 
We therefore conclude that our model is suitable for a study of the helium EUV resonance lines
under quiet sun conditions.

%%%%%%%%%%%%%%%
\subsection{Deriving the DEM from the model atmosphere}
%%%%%%%%%%%%%%%

%%%%%%%
\begin{table*}
\caption{Line ratios used to estimate the electron
  pressure.}
\label{tab:electronpressure}
\centering
\begin{tabular}{ccccc}
\hline \hline
    Lines & $^{10}\log{T_\mathrm{max}}$  & 
             Ratio & $n_\mathrm{e}$ &
            $n_\mathrm{e}T$ \\ 
    & (K) & & (10$^{10}$ cm$^{-3}$) & (10$^{15}$ cm$^{-3}$K) \\ \hline
  \ion{O}{v} 762.0 / 629.7    & 5.4 & 0.02  & 1.3 & 3.12 \\
  \ion{Si}{vii}  275.3 / 275.7   & 5.8  &   5.6 & 0.3 & 1.61  \\
  \ion{Fe}{xii} 193.5 / 186.9 & 6.2   &  1.9 & 0.1 & 1.67  \\ 
\hline
\end{tabular}
\end{table*}
%%%%%%%

%%%%%%%
\begin{figure}
 \includegraphics[width=\columnwidth]{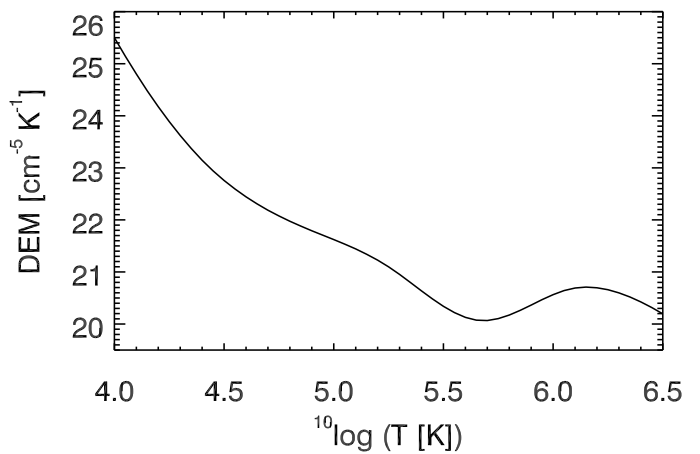}
   \caption{Differential emission measure derived from the model atmosphere. The DEM is
   derived from the intensities of the lines shown in Fig. \ref{fig:thin_giunta_comparison}}
 \label{fig:dem1d}
\end{figure}
%%%%%%

We follow \cite{2015ApJ...803...66G} and derive a DEM based on a spatial average of 
our computed line intensities, $\mean{I_\mathrm{th}}$. We use three density-sensitive 
line ratios to estimate the electron density at three different temperatures. The density 
sensitive line ratios and their values are listed in Table \ref{tab:electronpressure}. These are the same
line ratios as the ones used by \cite{2015ApJ...803...66G}. The electron
pressure varies less than the electron density over the temperature range. For this reason
we use the average electron pressure of $P_e/k_\mathrm{B}=n_\mathrm{e}T = 2\times10^{15}$ cm$^{-3}$~K
to derive the DEM. The line intensities we use to derive the DEM are the eleven  lines shown
in Figure \ref{fig:thin_giunta_comparison}. 
To perform the inversion we use the
XRT inversion code included in the Chianti package 
\citep{dere1997, 2015A&A...582A..56D}.
The resulting DEM is shown in Fig.~\ref{fig:dem1d}.
%

%%%%%%%%%%%%%%%%%
\subsection{Enhancement factors}
%%%%%%%%%%%%%%%%%

%%%%%%%
\begin{figure}
 \includegraphics[width=\columnwidth]{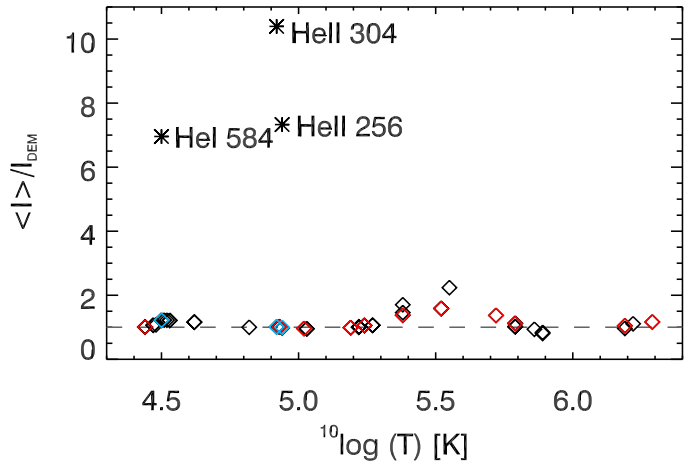}
   \caption{Enhancement factors for all the modelled lines. The lines selected for 
   DEM inversion are indicated by red diamonds. The helium lines are indicated
   by the black stars. The blue diamonds show the enhancement factors for the
   helium lines if we use their optically thin equilibrium intensity as the 'observation'.
   The black diamonds show the enhancement factor for all the other lines included in the study.}
 \label{fig:intrat1d}
\end{figure}
%%%%%%

We calculate the modelled line intensities, $I_\mathrm{DEM}$, with Eq.~\ref{eq:dem} assuming a
constant electron pressure (equal to the pressure used to derive the DEM). 
Figure \ref{fig:intrat1d} shows the synthetic enhancement factor, $\mean{I_\mathrm{th}}/I_\mathrm{DEM}$, for
all our included lines, except the helium lines where the enhancement factor is computed
by $\mean{I_\mathrm{RT}}/I_\mathrm{DEM}$. They are given as a function of formation temperature. 
Our DEM distribution successfully reproduces the observed intensities for most of the lines,
also those not included for the DEM inversion (indicated by the black diamonds). 
The helium lines stand out with enhancement
factors of about 7 for \ion{He}{i} $\lambda$584 and \ion{He}{ii} $\lambda$256, and 10 for
\ion{He}{ii} $\lambda$304. These values are comparable to values  
reported in earlier studies (see Table~\ref{table:efactors}).
We therefore consider the long standing problem of the anomalous helium line intensities in the quiet sun 
as solved.
The helium enhancement factors are caused by optically-thick radiative transfer effects and/or the 
effects of non-equilibrium ionisation.
In the next section we investigate which effects are important for the formation of \Hefef, \Hetfs, and \Hetof.

%%%%%%%%%%%%%%%%%%%%%%%%
\section{Relevant effects for the line intensities}\label{section:rteffects}
%%%%%%%%%%%%%%%%%%%%%%%%

%%%%%%%
\begin{table}
\caption{Ratios of average line intensities illustrating the effect
of thick radiative transfer.}
\label{table:lineratios}
\centering
\begin{tabular}{cccc}
\hline \hline
    Ratio & \ion{He}{i} $\lambda$584  & 
                  \ion{He}{ii} $\lambda$304 & \ion{He}{ii} $\lambda$256 \\ \hline
  $R_\mathrm{DEM}$ & 0.14 & 0.096 & 0.14 \\
  $R_\mathrm{th}$ & 0.18 & 0.097 & 0.14  \\
  $R_\mathrm{th,NE}$  & 0.097 & 0.31 & 0.61 \\
\hline
\end{tabular}
\end{table}
%%%%%%

We found in the previous section that the spatially averaged intensity, $\mean{I_\mathrm{RT}}$, is higher 
than what we obtain from the DEM distribution. Compared to the simple 1D DEM thin line formation,
we are including much more physics in the full NE-NLTE radiative transfer case. Among these ingredients
are multidimensional atmospheric structure and a non-equilibrium 
ionisation state. Ignoring radiative transfer effects, we begin this analysis by testing whether the enhanced intensity can be 
due to these two ingredients. 

First we asses how the 3D atmospheric structure alters the helium line intensity. 
The intensities computed by line-of-sight integration (Eq. \ref{eq:thinint}) assuming optically thin equilibrium conditions
include possible effects of such structure since the emissivity is calculated for each grid point
in the model atmosphere based on the local values of temperature and electron density.
To quantify the effect we 
define the line intensity ratios,
\begin{eqnarray}
R_{\mathrm{DEM}} &=& \frac{I_\mathrm{DEM}}{\mean{I_\mathrm{RT}}} \label{eq:rdem} \\
R_{\mathrm{th}}   &=& \frac{\mean{I_\mathrm{th}}}{\mean{I_\mathrm{RT}}}.  \label{eq:rth}
\end{eqnarray}
The values of the ratios for the helium lines are given in Table \ref{table:lineratios}. $R_\mathrm{DEM}$ 
is the inverse of the enhancement factor, and it has a value lower than 1 for the three helium lines considered.  
The values of $R_\mathrm{th}$ are very similar to the values of $R_\mathrm{DEM}$. This means that
the helium line intensities derived from the DEM approximately reproduce the spatial average
of the helium line intensities under optically thin equilibrium conditions. Or in other words, the 
multidimensional atmospheric structure does not explain the enhanced intensity.

Next we check the effect of non-equilibrium helium ionisation. To do this we compute
the optically thin non-equilibrium intensities, $I_\mathrm{th,NE}$. These intensities are 
calculated using the line-of-sight integration in Eq. \ref{eq:thinint}
of the thin non-equilibrium emissivity, $\eta_\mathrm{neq}$,  which features
the same excitation mechanism as the thin equilibrium emissivity, 
$(n_\mathrm{u}/n_\mathrm{ion})_\mathrm{eq}$, but we let the ion fraction be given by 
the non-equilibrium value from the model atmosphere, 
$(n_\mathrm{ion}/n_\mathrm{el})_\mathrm{neq}$. Then we can express the emissivity,
\begin{eqnarray}
  \eta_\mathrm{neq} = \frac{(n_\mathrm{ion}/n_\mathrm{el})_\mathrm{neq}} 
          {(n_\mathrm{ion}/n_\mathrm{el})_\mathrm{eq}} \ \eta_\mathrm{eq}
  \label{eq:eta_ne}
\end{eqnarray}
We are interested in how the spatial average of the thin non-equilibrium line intensity
differs from its equilibrium counterpart. We therefore define the ratio
\begin{eqnarray}
R_{\mathrm{th,NE}} = \frac{\mean{I_\mathrm{th,NE}}}{\mean{I_\mathrm{RT}}}.  
\label{eq:rthne}
\end{eqnarray}
The values of this ratio are given in Table \ref{table:lineratios}.

%%%%%%%%
\begin{figure*}
  \includegraphics[width=\textwidth]{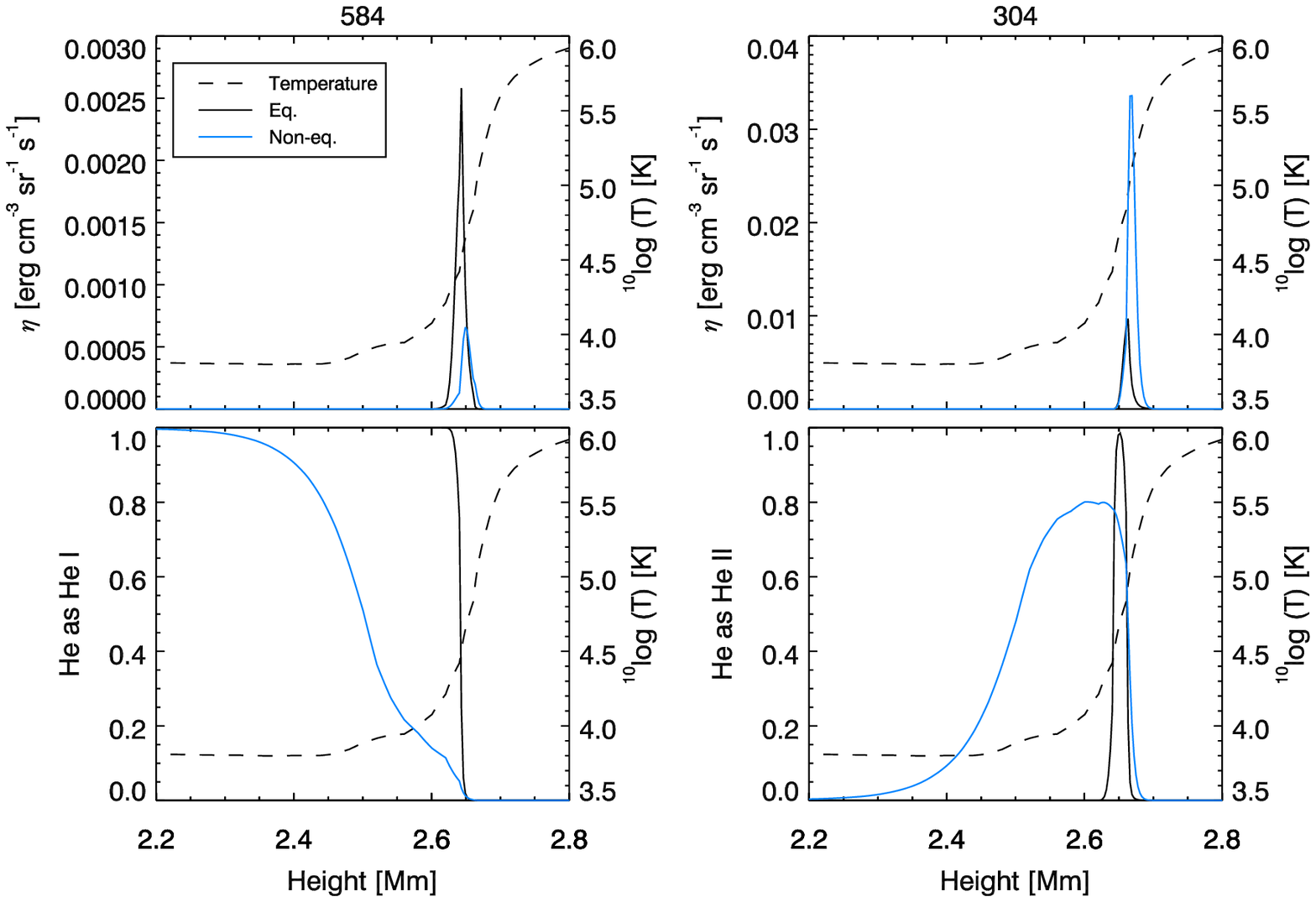}
  \caption{Details on the thin line formation of \ion{He}{i} $\lambda$584 (left column) and \ion{He}{ii} 
  $\lambda$304 (right column). Upper panels: line emissivities (solid, scale to the left) and temperature 
  (dashed, scale to the right). Bottom panels: ion fractions (solid, scale to the left) and temperature
  (dashed, scale to the right). The black lines show the values corresponding to ionisation equilibrium and the 
  blue lines show the values corresponding to non-equilibrium ionisation.}
  %JT the aspect ratio can be decreased (to the golden ratio for example (1+sqrt(5))/2 = 1.618). The axis labels appear too big.
  \label{fig:explain_thin}
\end{figure*}
%%%%%%%

Compared to $R_{\mathrm{th}}$, $R_\mathrm{th,NE}$ is halved for the \HeI\
line and increased by a factor of three for the \HeII\ lines. In other words, if the lines are only excited by collisions,
taking the non-equilibrium ionisation state into account results in fewer \Hefef\ photons and more \Hetof\ and \Hetfs\ photons 
than what is predicted by assuming ionisation equilibrium (as described in Section \ref{section:thinformation}). 
Figure \ref{fig:explain_thin} shows the ion fractions and corresponding emissivities
for the \Hefef\ and \Hetof\ lines in a network region column. We choose a network region column 
because the network regions feature the strongest emission and contribute the most to
the averages that goes into $R_\mathrm{th}$, and $R_\mathrm{th,NE}$.
We inspected a representative sample of columns in the snapshot, and they all behave qualitatively the same.
The non-equilibrium \HeI\ ion fraction is lower than the
ionisation equilibrium value, and this makes the non-equilibrium \Hefef\  emissivity lower.
It is opposite for the \Hetof\ emissivity;  the non-equilibrium \ion{He}{ii} ion fraction
has a tail that reaches into the high temperature transition region (at roughly $z=2.68$ Mm) which results in a 
non-equilibrium emissivity higher than the corresponding equilibrium value.

$I_\mathrm{RT}$ and $I_\mathrm{th,NE}$ are computed
from an identical atmosphere model and with identical ion fractions. 
Optically-thick radiative transfer effects therefore remain the only reason why $R_\mathrm{th,NE}$ is 
not equal to one. The ratio is closer to one for the \ion{He}{ii} lines than for the
\ion{He}{i} line. This suggests that the radiative transfer effects are more important for 
the latter. 

%%%%%%%%%%%%%%%%%%%%%%%%%%%%%%
\subsection{Radiative transfer effects on the He EUV lines}
%%%%%%%%%%%%%%%%%%%%%%%%%%%%%%

%%%%%%%
\begin{figure*}
  \includegraphics[width=\textwidth]{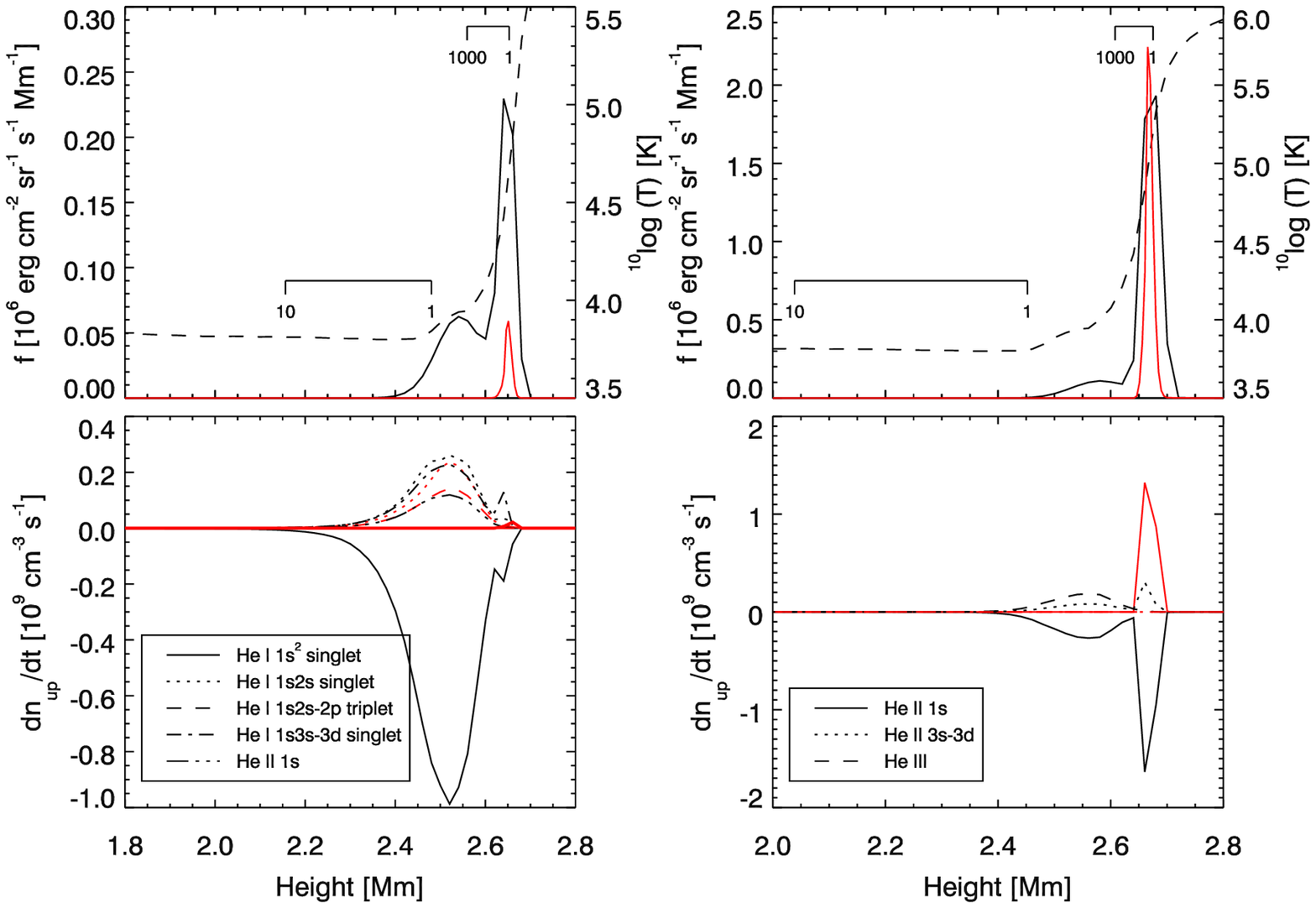}
  \caption{Comparison of the NE-NLTE thick and non-equilibrium thin line formation 
  of \ion{He}{i} $\lambda$584 (left column) and \ion{He}{ii} $\lambda$304 (right column).  
  Upper panels: contribution functions (scale to the left) for thick formation (solid black) and thin
  formation (solid red). For context the temperature is also shown (dashed curve, scale to the right). 
  Two additional scales are drawn in each of the two upper panels. The upper scale shows
  the optical depth at line center. The lower scale shows the optical depth
  for the continuum.
  Bottom panels: net rates into the upper level of the transition. Collisional processes are indicated
  by red lines and radiative transitions are indicated by black lines.}
  \label{fig:formation_584_304}
\end{figure*}
%%%%%%%

We now turn to the optically thick formation of the helium EUV line intensities. Fig.~\ref{fig:formation_584_304} shows the details  
for a network column (the same one as in Fig.~\ref{fig:explain_thin}).
Here we display a contribution function, $f(z)$.
It represents the contribution to the 
total line intensity as a function of height. We define it by using the formal solution
of Eq. \ref{eq:rt} \citep[page 38]{mihalas1978} to rewrite  
Eq.~\ref{eq:lineint},
\begin{eqnarray}
  I &=& \int_D I_\nu \ \dd\nu \nonumber \\
   &=& \int_D \int_{\tau_\nu} S_\nu e^{-\tau_\nu} \ \dd\tau_\nu \ \dd\nu \nonumber \\
   &=& \int_D \int_{z} \eta_\nu e^{-\tau_\nu} \ \dd z \ \dd\nu \nonumber \\
   &=& \int_z \int_D \eta_\nu e^{-\tau_\nu} \ \dd\nu \ \dd z \nonumber \\
   &=& \int_z f(z) \ \dd z \label{eq:contributionfunction}
\end{eqnarray}
The contribution function is equal to the emissivity (Eq.~\ref{eq:emissivity}) when the line formation is thin.

First we consider the \Hefef\ line.
The thick contribution function has a larger value than the thin contribution function
at all depths where it is significant. It is also more extended in space. The number of photons
released in the transition is closely related to the number density of atoms in the excited state
(Eq.~\ref{eq:emissivity}). 
In the lower panel we therefore show which processes populate the 
upper level of the transition in the thick calculation. Practically all of the transitions into the upper level are balanced by a radiative
de-excitation releasing a \Hefef\ photon. There is some collisional excitation in the transition region above $z=2.6$ Mm 
(which causes \emph{all} of the emission in the thin calculation). 
The dominant processes in the chromosphere below are radiative transitions from higher energy singlet states 
and collisional transitions from the triplet states. 
There is also a significant net radiative rate into the 
upper state from the lower energy singlet state 1s2s. The singlet state 1s2s is populated mainly by 
collisions from the triplet states (not shown in figure). In the chromosphere the triplet states are populated 
mainly by recombinations from \ion{He}{ii} 1s 
\citep[e.g.,][]{andretta1997, centeno2008}. 
This means that a significant
amount of the \Hefef\ photons are due to recombination from singly ionised helium followed by a cascade down to the 
upper level of the line transition, either through the singlet system or the triplet system. 
Recombination into excited states is not included in thin modelling. A DEM analysis would therefore 
never account for these photons.

%%%%%%%
\begin{figure}
  \includegraphics[width=\columnwidth]{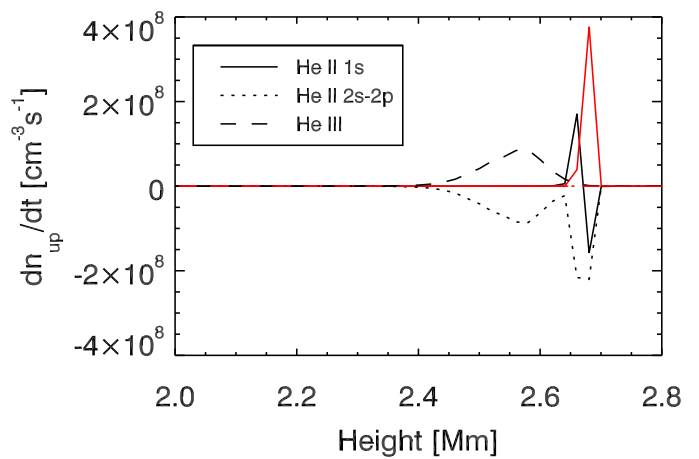}
  \caption{Net rates into upper level of the \ion{He}{ii} $\lambda$256 line transition. The 
  red lines indicate collisional transitions. The black lines indicate radiative transitions.}
  \label{fig:contribution256}
\end{figure}
%%%%%%

Next we consider the \Hetof\ line. The thick contribution function is more extended in height than the thin 
contribution. But both the thin and thick contribution functions peak in the transition region.
The thin $f$ has a higher peak than the thick $f$ because it is not depressed by optical depth effects.
The dominant process populating the upper level of the transition where the contribution function is significant, is
collisional excitation from the \ion{He}{ii} ground state. There is a smaller contribution to the total \Hetof\ 
line intensity that comes from the chromosphere. Here recombination of \ion{He}{iii} dominates 
in populating the excited state.

For the \Hetof\ line there are two radiative transfer effects at play: The first one is 
backscattering of photons. When a collisionnally excited \ion{He}{ii} ion de-excites, it releases
a photon that  travels either up into space or down towards the chromosphere. The line opacity
in the chromosphere is high and the photon destruction probability is low. The photon will thus be scattered around. Since the destruction probability is low, a large fraction of the downward-emitted photons will eventually escape upwards into space.

Under idealised conditions, where the photon production happens in a truly thin layer and the destruction probability is zero, all photons eventually escape, and therefore $\mean{I_\mathrm{RT}} = 2 \mean{I_\mathrm{th,NE}}$ and $R_\mathrm{th,NE} = 0.5$ . Under realistic conditions, the proportionality constant is smaller than two.

The calculated value of $R_\mathrm{th,NE}$ is lower than 0.5, or equivalently $\mean{I_\mathrm{RT}} > 2 \mean{I_\mathrm{th,NE}}$, which means that backscattering alone is not sufficient.
We explain this by the second effect which is the recombination from \ion{He}{iii}. The last step of 
the recombination cascade produces a \Hetof\ photon. These photons are primarily created in the optically 
thick chromosphere. They are released into space after a sequence of scattering steps and add 
to the thick line intensity.

Turning finally to the \ion{He}{ii} $\lambda$256 we have $R_\mathrm{th,NE} =0.61$. This line is also optically thick in the chromosphere so we expect that the backscattering
effect is important. Since $R_\mathrm{th,NE}$ is significantly higher than 0.5, a fraction of the line photons 
must be destroyed. Figure \ref{fig:contribution256} shows the net rates into to excited state of the line
in the same network column as shown in Figure \ref{fig:formation_584_304}. Indeed, \Hetfs\ photons
are absorbed and transformed into \ion{He}{ii} Balmer-$\alpha$ (and later 304) photons.
The dominant processes populating the upper level of the line are otherwise very similar to 
those populating the excited state of the \Hetof\ line, with the recombination of \ion{He}{iii} in the 
chromosphere. However, this recombination is the beginning of the downward cascade 
and the production of a \ion{He}{ii} Balmer-$\alpha$ photon and a \Hetof\ photon. 

To sum up we find that the \ion{He}{i} $\lambda$584 line is in large part formed by 
recombination cascades and that the resonance lines of \ion{He}{ii} are mostly 
collisionally excited. We have found this based on radiative transfer modelling 
where we have taken time dependent ionisation driven by coronal illumination into account.
Next we evaluate the effect of how time dependence and coronal illumination
affect the intensities of the helium EUV resonance lines.

%%%%%%%%%%%%%%%%%%
\subsection{Non-equilibrium effects}\label{section:noneqeffects}
%%%%%%%%%%%%%%%%%%

%%%%%%%
\begin{table}
\caption{Ratios of average line intensities illustrating the effect of non-equilibrium 
           ionisation.}
\label{table:lineratios_timedep}
\centering
\begin{tabular}{cccc}
\hline \hline
    Ratio & \ion{He}{i} $\lambda$584  & 
                  \ion{He}{ii} $\lambda$304 & \ion{He}{ii} $\lambda$256 \\ \hline
  $R_\mathrm{RTE}$ & 0.88 & 0.52 & 0.57 \\
%  $R_\mathrm{RT}$ & 1.0 & 1.0 & 1.0  \\
\hline
\end{tabular}
\end{table}
%%%%%%

%%%%%%%
\begin{figure}
  \includegraphics[width=\columnwidth]{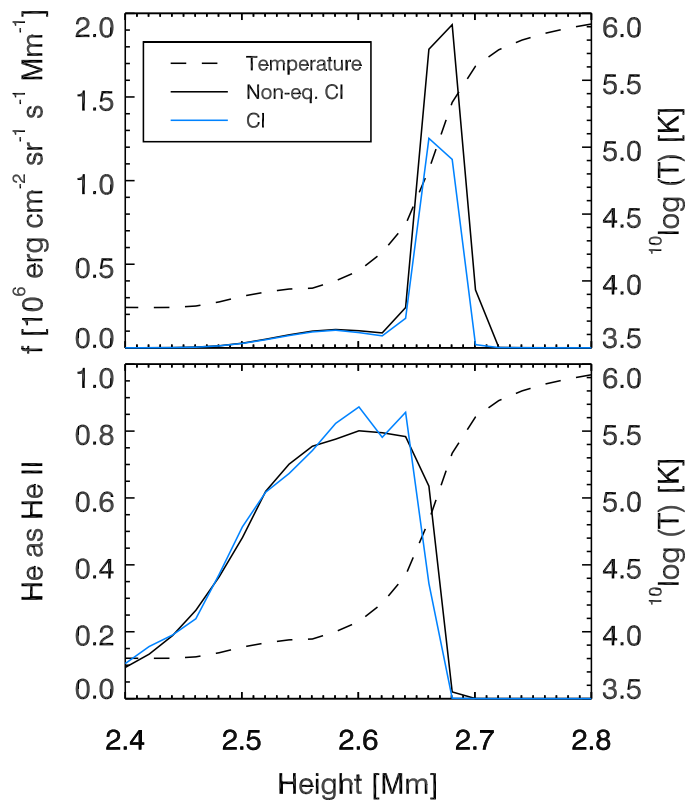}
  \caption{Details on the \ion{He}{ii} $\lambda$304 thick line formation under 
  equilibrium and non-equilibrium conditions. Upper panel: contribution 
  functions (solid lines, scale to the left) and temperature (dashed line, 
  scales to the right). Lower panel: Fraction of He as \ion{He}{ii} (solid 
  lines, scale to the left) and temperature (dashed line, scales to the
  right).}
  \label{fig:timedep_he2}
\end{figure}
%%%%%%

To identify the non-equilibrium effects inherent in $I_{\mathrm{RT}}$ we compute 
an equilibrium solution $I_\mathrm{RTE}$ and compare the two. The equilibrium
solution is computed by using equilibrium ion fractions as constraints in the NE-NLTE mode
of Multi3d. The equilibrium ion fraction are computed on the basis of the
coronal illumination present in the Bifrost simulation. The simulation uses a 3-level helium
model atom. To advance the atomic populations, Bifrost computes the radiative and collisional 
ionisation and recombination transition rate coefficients for every time step. The radiative
ionisation coefficient indirectly contains the coronal illumination. We obtain the equilibrium ion fractions by 
using the radiative and collisional transition coefficients from the Bifrost simulation 
to solve two equilibrium rate equations (eq. \ref{eq:rate}) and one particle conservation equation
(eq. \ref{eq:partcons}). This way both $I_\mathrm{RT}$ and $I_\mathrm{RTE}$ include
the effects of coronal illumination, but all non-equilibrium effects are removed from $I_\mathrm{RTE}$.
The equilibrium line intensities,  $I_\mathrm{RTE}$, are in principle comparable to 
line intensities computed from a radiative transfer statistical equilibrium setup where coronal illumination
is taken into account, either as an incoming EUV radiation field in the upper boundary 
\citep{avrett1994, 
wahlstrom1994, andretta1997, pietarila2004, mauas2005, centeno2008},
or as an extra emissivity included in the calculations
 \citep{leenaarts2016}.

Table \ref{table:lineratios_timedep} shows the ratios of the spatially averaged 
line intensities,
\begin{eqnarray}
  R_\mathrm{RTE}&=& \frac{\mean{I_\mathrm{RTE}}}{\mean{I_\mathrm{RT}}}  
%  R_\mathrm{RT}&=& \frac{<I_\mathrm{RT}>}{<I_\mathrm{RT}>}.
\end{eqnarray}
Roughly a tenth of the \ion{He}{i} $\lambda$584 photons and
half of the \ion{He}{ii} $\lambda$304 and $\lambda$256 photons can be attributed to
non-equilibrium effects. Figure~\ref{fig:timedep_he2} shows the difference of
equilibrium and non-equilibrium \ion{He}{ii} $\lambda$304 line formation in the same column
as the column featured in Fig.~\ref{fig:formation_584_304}.
In the transition region, where the line is collissionally excited, the non-equilibrium contribution 
function is higher than the equilibrium contribution function. This is consistent
with the differences in the \ion{He}{ii} fraction. In the high temperature transition region the
equilibrium ion fraction is lower than the non-equilibrium ion fraction. 
In other words, time dependent ionisation permits 
\ion{He}{ii} to exist in a hotter environment than what is permitted under equilibrium 
conditions. The result is more collisional excitation and a higher
line intensity. This applies to both of the \ion{He}{ii} lines.

%%%%%%%%%%%%%%%%%%%%%
\subsection{Effects of coronal illumination}
%%%%%%%%%%%%%%%%%%%%%

%%%%%%%
\begin{figure}
  \includegraphics[width=\columnwidth]{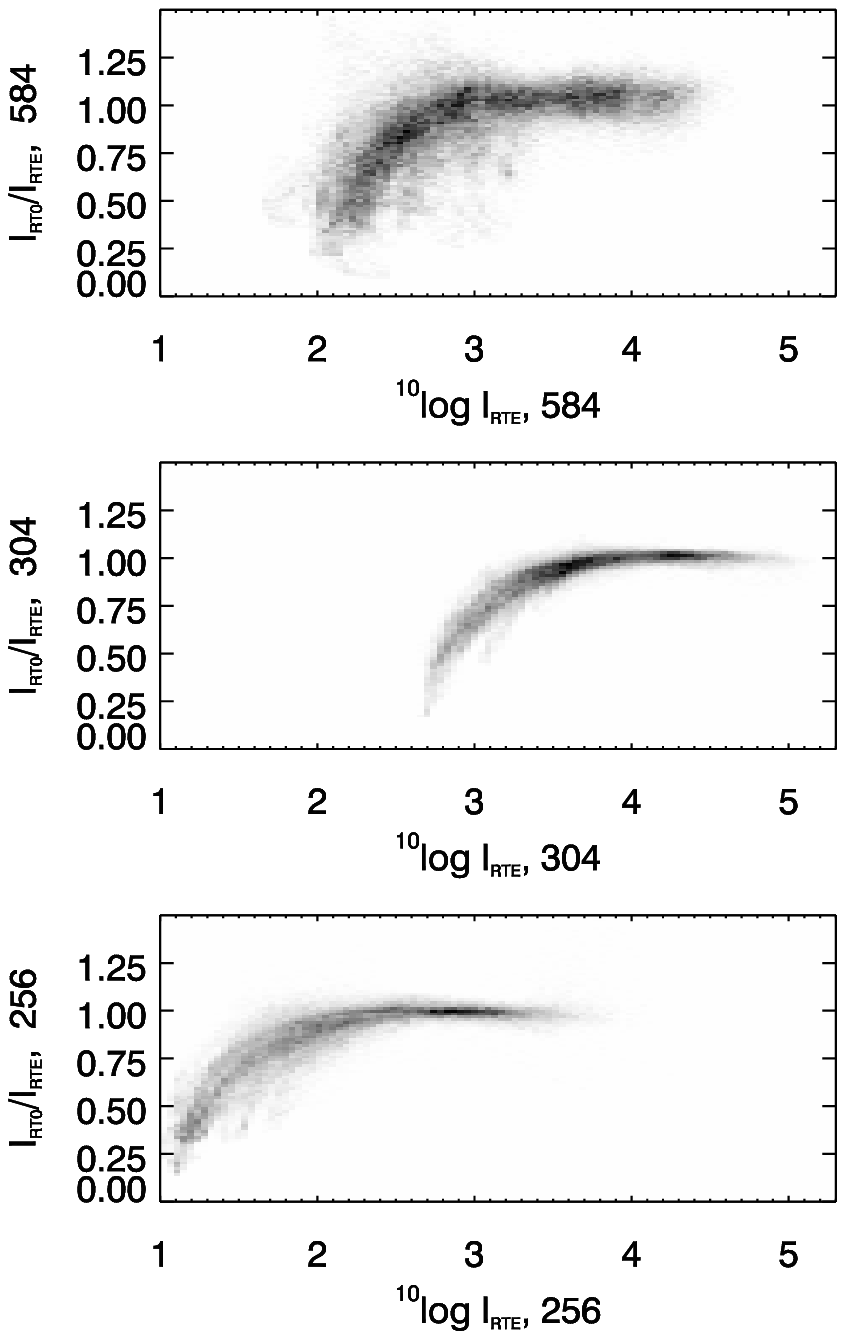}
  \caption{Joint probabilty distributions of the ratio $R_\mathrm{CI} = 
  I_\mathrm{RT0}/I_\mathrm{RTE}$ as
  a function of $I_\mathrm{RTE}$. Tops to bottom panels show the relations
  for \Hefef, \Hetof, and \Hetfs.}
  \label{fig:noci_ci}
\end{figure}
%%%%%%
%%%%%%%
\begin{table}
\caption{Ratios of average line intensities illustrating the effect of coronal illumination.}
\label{table:lineratios_ci}
\centering
\begin{tabular}{cccc}
\hline \hline
   Ratio & \ion{He}{i} $\lambda$584  & 
                  \ion{He}{ii} $\lambda$304 & \ion{He}{ii} $\lambda$256 \\ \hline
 $R_\mathrm{CI}$ & 1.01 & 0.98 & 0.97 \\
\hline
\end{tabular}
\end{table}
%%%%%%

Next we assess the effects of coronal illumination. 
To remove this effect, and this effect alone, we need to run a new Bifrost simulation including 
non-equilibrium ionisation but with the photoionisation switched off, and with a snapshot from this
simulation calculate the NE-NLTE radiative transfer solution with the new ion fractions.  
This costs more than we can afford. Instead we compare the equilibrium line intensities, $I_\mathrm{RTE}$, 
with non-LTE line intensities, $I_\mathrm{RT0}$. The former includes the effect
of coronal illumination, the latter does not.

Table \ref{table:lineratios_ci} gives the ratio  
\begin{eqnarray}
  R_\mathrm{CI} = \frac{\mean{I_\mathrm{RT0}}}{\mean{I_\mathrm{RTE}}} 
\end{eqnarray}
for the three resonance lines. The value is approximately one for all of them.
This suggests that the effect of coronal illumination is negligible for the {\em mean} quantities.
\cite{andretta1997} found that the effect of an incident EUV radiation field influenced the line
intensity of \Hefef\ only for their models featuring a low density 
transition region. Their models with a high density transition region had a low sensitivity 
to the coronal illumination. The high intensity network regions in our model correspond to the high density 
transition region models of \cite{andretta1997}. These regions dominate in the line intensity averages, and 
this explains why the values of $R_\mathrm{CI}$ are approximately equal to one. 
To verify this explanation
we visualise the effect of coronal illumination for all columns of our model in Figure~\ref{fig:noci_ci}, which
shows the distribution of $I_\mathrm{RT0}/I_\mathrm{RTE}$ as a function of $I_\mathrm{RTE}$. For the \Hefef\ 
line the ratio is lower than one in low intensity regions and approximately one in high intensity regions. 
This is the trend also for the \Hetof\ and \Hetfs\ lines showing that the effect of the coronal illumination
is more important in low intensity regions.

%%%%%%%%%%%
\section{Discussion}\label{section:discussion}
%%%%%%%%%%

%%%%%%%
\begin{figure}
  \includegraphics[width=\columnwidth]{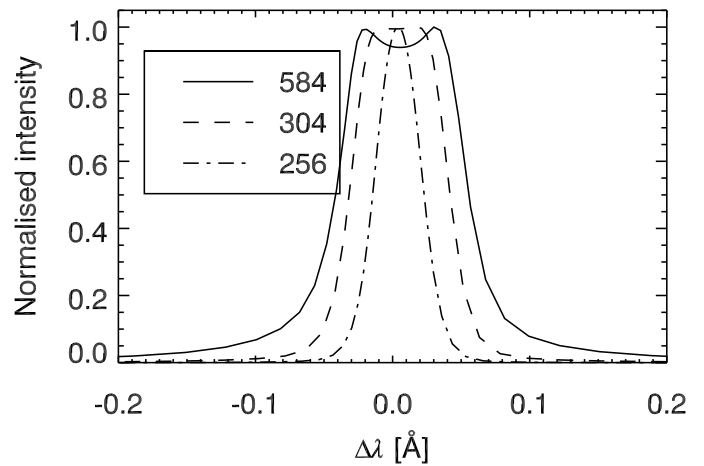}
  \caption{Average line profiles based on the NE-NLTE intensity of the whole model atmosphere.}
  \label{fig:profiles}
\end{figure}
%%%%%%%

Suggestions on various enhancement mechanisms have been made in the past.
Particle diffusion \citep{falc1993} and velocity redistribution \citep{andretta2000} are examples 
that would both lead to a mixing of hot electrons and cold ions.
Presence of a macroscopic velocity field across the transition region will
have a qualitatively similar effect, as was shown by 
\citet{fontenla2002} 
in static 1D models. 
Our model atmosphere and radiative transfer calculations take time dependent 3D flows
and non-equilibrium ionization into account, and we end up with another
scenario of hot electrons exciting cold atoms (shown in Fig.~\ref{fig:timedep_he2}). 
While we will not rule out possible effects due to particle diffusion and velocity
redistribution,
we can safely say that non-equilibrium 
effects enhance the intensities.

The calculated intensities are dependent on the ionization state, and the ionization
state is dependent on the 3D structure and 3D radiation of the atmosphere. Modelling the 
radiation field in 1D will potentially lead to erroneous ion fractions and therefore erroneous 
intensities. For instance, a chromospheric 
structure may be exposed to EUV radiation from the sides as well as from the top. Modelling the 
the radiation in 1D would expose such a structure to EUV radiation only from the top which
would result in a lower photoionization rate and fewer ions. An example of this is given in 
Figure 7 of \cite{leenaarts2016} where
a comparison between 1D and 3D radiative transfer shows that an
exposed chromospheric structure has a higher number density of \HeII\ in the 3D case than
in the 1D case.

Another enhancement mechanism that has been suggested is the PR-mechanism. 
We have shown that a significant amount of the \ion{He}{i} $\lambda$584 photons are created in 
recombination cascades. This is valid in both network and internetwork regions of our model.
According to 
\cite{1975ApJ...199L.131M} 
the PR-mechanism should lead to a central 
reversal in the line profile. Figure~\ref{fig:profiles} shows the average profiles of all three of the studied 
resonance lines, where the intensity is from NE-NLTE radiative transfer calculation.
The \Hefef\ profile clearly has a central reversal, even after averaging over all the
columns. We found that the coronal illumination 
is important for the 584 line intensity in the low-intensity internetwork regions, but less so in the high 
intensity network regions. The column  shown in Figure \ref{fig:formation_584_304} is from such a high intensity 
network column, and the figure shows the effect of recombination cascades.  This 
suggests that the photoionisation in the network is dominated by internal sources of EUV photons - 
not an externally set coronal illumination. 
The possible internal sources of photons capable of ionising \ion{He}{i} included in the calculations 
are \Hetof, \Hetfs, and the two helium continua. Also photons from the hydrogen Lyman continuum 
shortward of 504 \AA\ are capable of ionising \ion{He}{i}. The hydrogen Lyman continuum is included as a background element,
and the contribution is determined by the non-equilibrium hydrogen populations from the Bifrost
simulation. An investigation of the relative
importance of these internal photon sources is beyond the scope of the present study.
 
These findings are consistent with \cite{andretta1997}, 
who found that the resonance line of neutral helium has a low sensitivity to coronal illumination in models 
with a high density transition region. The question is then: have \Hefef\ central reversals been observed?
A central reversal was measured by 
\cite{1982JGR....87.1433P}.
 \cite{1997SoPh..170...75W}
reported single peaked average profiles based on observations from SOHO/SUMER 
\citep{1995SoPh..162..189W}.
Later both 
\cite{1999ApJ...522L..77P} 
and 
\cite{2004ApJ...606.1258J} 
found indications of central reversals, also based on observations from SOHO/SUMER. The line was 
also observed with SOHO/CDS  
\citep{1995SoPh..162..233H},
but since the CDS instrument response function is broad, any double-peaked \Hefef\ profiles would have 
been washed out as was shown by
 \cite{mauas2005}.
Our results can be brought into doubt if central reversals are uncommon features for the \Hefef\ profile.
They have been observed, but not everywhere. However, the observation of narrow double-peaked profiles may
have been prevented by a too low spectral resolution of the relevant instruments. We believe therefore
that the central reversal present in our modelled \Hefef\ profiles is not in conflict with observations. 
However, further studies on this issue are merited.

We found that the resonance lines of \ion{He}{ii} are primarily excited by collisions. The role of PR
is moderate for \Hetof\ and marginal for \Hetfs. This is also reflected in the average profiles 
shown in figure \ref{fig:profiles}: the \Hetof\ profile has a slightly flattened peak and the \Hetfs\ profile appears
Gaussian. Our finding that the \ion{He}{ii} lines are collisionnally excited in the quiet sun is in 
agreement with what has been concluded in earlier studies 
\citep{1993ApJ...406..346J, andretta2003,
2007ASPC..368..183J}.
%

%%%%%%%%%%%%%%%%%%
\section{Summary and conclusion}\label{section:summary}
%%%%%%%%%%%%%%%%%%%

We use the code  Multi3d to solve the radiation transfer problem for helium on a
radiation-MHD simulation snapshot. By constraining the radiative transfer 
solution with the non-equilibrium helium ion fractions, we are able to obtain non-equilibrium 
non-LTE atomic spectra. The EUV resonance lines \Hefef, \Hetof, and \Hetfs\ have intensities 
an order of magnitude higher than what we get by assuming optically thin equilibrium conditions.
This difference is in line with the difference between observed and modelled  intensities
reported in the literature.

The high NE-NLTE intensity of \Hefef\ is explained by large contributions from recombination
cascades. The high NE-NLTE intensity of \Hetof\ and \Hetfs\ is explained primarily by non-equilibrium 
effects. A higher number of \HeII\ ions are present in the high temperature transition region that what is 
permitted under equilibrium conditions. The collisional excitation rates of the \HeII\ line transitions are 
increasing with temperature. Thus there is more collisional excitation and more photons are produced in the 
lines when non-equilibrium ionisation is taken into account. 

We conclude that the problem of the anomalously high helium EUV line intensities disappears when taking into account optically thick radiative transfer and non-equilibrium ionisation effects.

\begin{acknowledgements} 
This research was supported by the Research Council of Norway through the 
grant ``Solar Atmospheric Modelling'', through grants of computing time from the
Programme for Supercomputing,  
and by the European Research Council under the European Union's 
Seventh Framework Programme (FP7/2007-2013) / ERC Grant 
agreement nr. 291058.  Some computations were performed on resources provided by the Swedish National
  Infrastructure for Computing (SNIC) at PDC Centre for High Performance Computing (PDC-HPC)
  at the Royal Institute of Technology in Stockholm. 
\end{acknowledgements}

\bibliographystyle{aa}
\bibliography{references}

\end{document}